\documentclass[12pt]{article}

\usepackage{epsfig}

\begin{document}
 
\sloppy


\baselineskip=6.5mm
\parskip=1.5mm

\begin{flushright}            
DCPT-02/90\\
IPPP-02/45\\                        
\end{flushright}
\begin{center}
{\Large{\bf TRANSLATING QUARK DYNAMICS \\INTO HADRON PHYSICS (and back again)}}\footnote{Presented at MESON2002,
Workshop on Production, Properties and Interactions of Mesons,\\               ~~~~~~~~~~~Cracow, Poland, May 2002}
\\[5mm]

{\large{\bf M. R. PENNINGTON}}\\[5mm]
{Institute for Particle Physics Phenomenology,\\ University of Durham, Durham DH1 3LE, U.K.}\\[5mm]

\large{Abstract}\\[2mm]
\end{center}

\baselineskip=5mm
{\leftskip 7mm\rightskip 7mm{Why should we study mesons in 2002? 
Two approaches to relating quark and gluon dynamics to hadron physics, 
namely QCD sum rules and effective field theories, are briefly discussed. 
These are linked by progress in the study of strong QCD, 
both by lattice Monte Carlo methods and in the continuum. 
These provide the translation from quark dynamics to hadron physics and 
back again. Scope for more theoretical work and further experiments to 
elucidate the nature of the QCD vacuum
and its precise relation to scalar mesons and their interactions is outlined.}}

PACS~~{13.75.-n, 13.65.+i, 13.25.-k, 14.40.Gx}

\baselineskip=6.mm
\section{MESONS in 2002?}

Why in 2002 should we be interested in low energy meson physics? After all, it is forty years since Gell-Mann~[1] recognised the $SU(3)$ flavour structure of baryons and mesons, and shortly after realised~[2] that this was naturally embodied in the quark model. All the evidence since then has confirmed that hadrons are indeed made of quarks. Moreover, it is now thirty  years since it was acknowledged that the force that binds quarks to make hadrons is mediated by gluons coupling to colour charges and described by the Lagrangian of QCD~[3] --- a seemingly simple equation. However, despite the lapse of thirty years, we do not yet know how to translate the hieroglyphs of the {\it Book of QCD} into the \lq\lq hadron text'' describing what we see in experiment (and vice versa). At first this is surprising since QCD is modelled on our best tested theory, 
namely QED, which makes very precise predictions, for instance, about the structure of atoms. These predictions rely on perturbation theory, in which an atomic electron, for example, moves through empty space. This we call the {\it vacuum}. Perturbations about this vacuum make it not quite empty. As the electrons orbit they radiate photons and these in turn can produce $e^+e^-$ pairs. As a result each electron swims in a (dilute) sea of fermion-antifermion pairs, the effect of which can be readily calculated.
At first we expect the vacuum of QCD to be very similar. Quarks move
through a cloud of gluons and a sea of $q{\overline q}$ pairs. Indeed, the effect of the gluon cloud is to make the interaction of the quarks weak at short distances. This \lq\lq asymptotic freedom'' makes perturbation theory most reliable for short distance interactions.

The simplest process producing quarks is $e^+e^-$ annihilation into hadrons. There at SLAC or LEP an electron and a positron annihilate in a region of space no more than a hundredth of the size of a proton. They create a quark and an antiquark which eventually produce jets of hadrons. To predict the resulting cross-section, we need not know how the quarks become hadrons, only that they do
so with unit probability. Thus using perturbative QCD we can predict the behaviour for $e^+e^-\to\; hadrons$ for each flavour of quark above the corresponding threshold. On their way to hadronisation the quarks propagate through the QCD vacuum containing $q{\overline q}$ pairs and a gluon cloud. Over longer distances
these quarks and antiquarks and gluons are so strongly interacting they form quark and gluon condensates. This alters the nature of the vacuum dramatically.
Fortunately, we can learn about this structure of the QCD vacuum in several ways which we will now discuss.

\section{QCD Sum Rules}

The first assumes a hadron description takes place at low energies (for light flavours) and a calculable QCD component at higher energies. This is embodied in the QCD sum rules proposed by Shifman, Vainshtein and Zakharov~[4] 25 years ago. We consider a correlator of two currents. The simplest example is the electromagnetic current of $e^+e^-$ annihilation. The correlator depends on $s$ the square of the momentum flowing in the current giving $\Pi(s)$. At low energies, this correlator is dominated by resonances, $\rho$, $\omega$, etc., and at high energies by
$u{\overline u}$ and $d{\overline d}$ loops corrected by gluon emission and the effect of higher dimension operators that are the condensates. $\Pi(s)$ is an analytic function in the complex $s-$plane with a cut along the real axis,
starting at the lowest hadronic threshold. Along the top of this cut is where experiments are performed. Since $\Pi(s)$ is an analytic function, it satisfies Cauchy's theorem round any closed contour in the complex $s-$plane:
\begin{equation}
\oint_C\, ds\; \omega(s)\,\Pi(s)\;=\;0
\end{equation}
provided the contour $C$ encircles no poles of $\Pi(s)$.
\begin{figure}[h]
\begin{center}
\mbox{~\epsfig{file=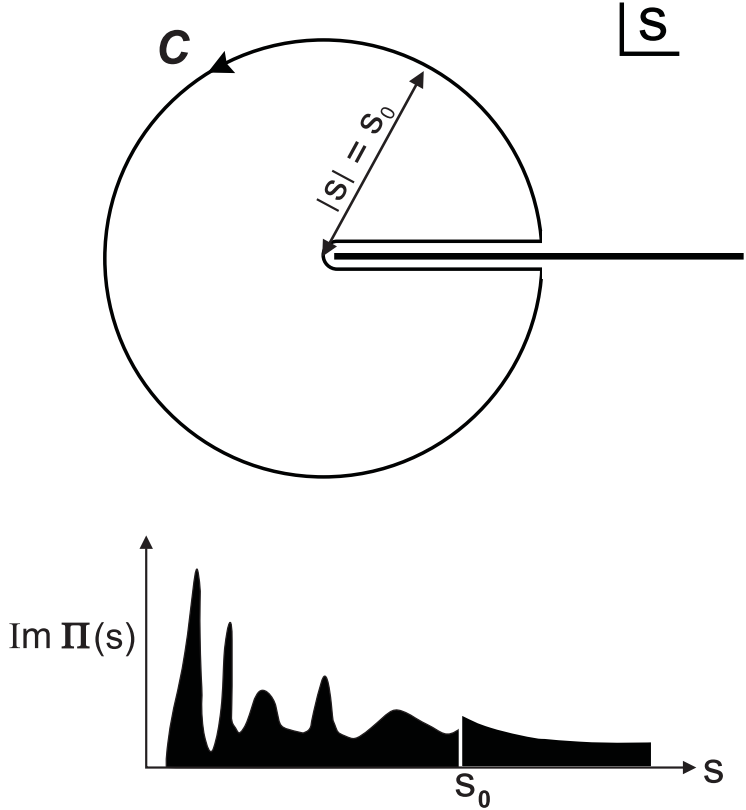,angle=0,width=9.8cm}}
\end{center}
\vspace{-1mm}
\caption{\leftskip 5mm\rightskip 5mm{Contour in the complex $s-$plane used for finite energy sum rules.
The lower figure illustrates how the contribution to the discontinuity across the cut is given by resonance physics at low energies,
while above $s=s_0$ it is calculable using the Operator Product Expansion from QCD. In reality, for finite energy sum rules the latter contribution is calculated round the circle.
Nevertheless, there is in general a mismatch at $s=s_0$ between the low energy hadronic component  and the higher energy pQCD component of the sum-rule integrand, as shown in the lower figure. }}
\end{figure}
\begin{table}[t]
\begin{center}
\begin{tabular}{|c|c|c|}
\hline
& & $\hat m(2~\textrm{GeV})~(\textrm{MeV})$\\ 
\hline
Chetyrkin \emph{et al.} [7] & Pseudoscalar & $4.0 \pm 0.8$\\
Prades [8] & Sum Rules & $4.5 \pm 0.9$\\
Maltman \& Kambor [9] & & $3.9 \pm 0.6$\\
&&\\
Cherry \& Pennington [10] & Scalar Sum Rules & $3.7 \pm 0.4$\\
&&\\
JLQCD [11] & Quenched & $4.2 \pm 0.3$\\
QCDSF [12] & Lattice QCD & $4.4 \pm 0.2$\\
APE [13] & & $4.8 \pm 0.5$\\
CP-PACS [14] & & $4.4 \pm 0.1$\\
&&\\
SESAM [15] & Unquenched & $2.7 \pm 0.1$\\
CP-PACS [14] & Lattice QCD & $3.5 \pm 0.2$\\
QCDSF - UKQCD [16] & & $3.5 \pm 0.2$\\
SESAM [17] & & $4.5 \pm 1.7$\\
\hline
\end{tabular}
\end{center}
\vspace{-1.5mm}
\caption{\leftskip 5mm\rightskip 5mm{The mean $u$ and $d$ quark mass at a scale of 2 GeV from sum rule analyses amd lattice computations, both quenched and unquenched.}}
\end{table}

 Here $\omega(s)$ is some suitable weight function. By taking the contour, $C$, as in Fig.~1, we can input experimental information of the appropriate quantum numbers along the cut up to $s=s_0$. Then round the contour $|s|\,=\,s_0$, we assume the correlator is describable by the part of QCD explicitly calculable using the Operator Product Expansion --- perturbation theory plus condensates (and instantons where appropriate). Though such sum rules have been in use for 25 years, developments in the past ten years have transformed their use from an art to a science, aided by the calculation of higher order perturbative contributions needed for the particularly useful finite energy sum-rules~[5]. Moreover, we have understood how to suppress the sensitivity to the transition illustrated in Fig.~1  from hadron physics to quark dynamics at $|s| = s_0$, by introducing so called \lq\lq pinched weights'' with a zero in the weight function, $\omega(s)$, at $s = s_0$~[6]. With these advances a series of sum rule analyses have been used to learn about the key parameters of QCD from hadron physics. In Table~1 are listed recent results~[7-17] for the mean {\it up} and {\it down} quark mass, $\hat m$, at the perturbative scale of 2 GeV in the ${\overline{MS}}$ scheme.
These show $\hat m\,\simeq\,3-5$ MeV. 

Similarly from Table 2 we see the
$q{\overline q}$ condensate ($= \langle u{\overline u}\rangle\,=\,\langle d{\overline d}\rangle$)~[18-22] has a scale of $-$(250-270 MeV)$^3$, again at a  \lq\lq perturbative'' scale  of 2 GeV. A recent major industry has been the extraction of the strange quark mass, which is between 90 and 130 MeV at a scale of 2 GeV~[23]. The corresponding
$\langle s{\overline s}\rangle$-condensate is however rather poorly determined~[18,24]. We will see these key numbers again.

\begin{table} [b]
\begin{center}
\begin{tabular}{|c|c|c|}
\hline
&& $\langle \bar{q} q \rangle (2\textrm{GeV})$\\
\hline
Narison [18] & Pseudoscalar Sum Rules & $-(247 \pm 9 ~ \textrm{MeV})^3$\\
&&\\
Dosch \& Narison [19] & $D$-decay Sum Rules & $-(212 - 289~\textrm{MeV})^3$\\ 
&&\\
Giusti \emph{et al} [20] & Quenched & $-(245 \pm 12 ~\textrm{MeV})^3$\\
Hern\'andez \emph {et al} [21] & Lattice QCD & $-(278 \pm 12~\textrm{MeV})^3$\\
MILC [22] & & $-(290 \pm 6~\textrm{MeV})^3$\\
\hline
\end{tabular}
\caption{\leftskip 5mm\rightskip 5mm{Values for the $u{\overline u}$, $d{\overline d}$ condensate
in the ${\overline{MS}}$ scheme at a scale of 2 GeV from sum rule and lattice analyses.}}
\end{center}
\end{table}

\section{Effective Lagrangians}
\baselineskip=6mm

An alternative way to relate quark dynamics to hadron physics is the method of effective field theories~[25]. At short distances interactions are described in terms of QCD with its renormalizable Lagrangian with a small number of basic interactions of quarks and gluons and a single coupling. At larger distances we imagine a description in terms of an effective Lagrangian of hadron interactions. In this all permitted couplings occur. There  are a huge number of terms. The theory is not renormalizable in terms of a finite number of constants.
Even at some finite order in the number of interactions 
very many are required. All have to be fixed from experiment before predictions can be made. Nevertheless, this approach has value in limited energy
regimes for particular processes. This comes about because we expect the effective 
hadronic Lagrangian to incorporate the symmetries of the underlying theory. So, for example, as a result of the near equality of the {\it up} and {\it down} quark masses, the QCD Lagrangian is symmetric in the  $u$ and $d$ quark fields. We correspondingly expect the hadronic Lagrangian to respect this $SU(2)$ symmetry (and even an approximate $SU(3)$ symmetry), so that protons and neutrons have the same strong interaction. 

Now the mass that enters the renormalized Lagrangian of QCD is the current mass, which for {\it up} and {\it down} quarks is only a few MeV, as we have seen, and very much less than the scale of $\Lambda_{QCD}$ of 100-200 MeV. Consequently, to a good approximation we can regard the {\it up} and {\it down} quarks as massless.
This means that their two helicity components become independent of each other.
We can therefore interchange {\it up} and {\it down} quark fields spinning left-handedly quite independently of the right-handed sector, and vice-versa. Consequently the QCD Lagrangian has an $SU(2) \otimes SU(2)$ chiral symmetry, which the hadronic Lagrangian should also respect. This clearly restricts the form of the allowed interactions~[26,27].  

Now this symmetry is not observed in the hadron world. Scalars and pseudoscalars, vectors and axial-vectors are not degenerate in mass with simply related interactions. This symmetry must be spontaneously broken. A mechanism that achieves this was proposed 40 years ago by Nambu~[28], long before the discovery of QCD. To see this, consider an effective Lagrangian of pseudoscalar ${\underline \pi}$ and scalar $\sigma$ fields. The potential generated by their interactions has to be
symmetric between these fields. If instead of this potential having the shape of a parabolic bowl with its minimum at zero values of the ${\underline \pi}$ and $\sigma$ fields, it has a Mexican hat (or wine bottle) shape, then the ground state chosen by nature has a non-zero value for the $\sigma$ field. The fluctuations about this minimum corresponding to the scalar particle are up and down the sides of the potential. This makes the $\sigma$ massive. Indeed, this
makes this scalar field the Higgs boson of the strong interactions. Its mass gives mass to all other hadrons, as we see below.
In contrast, the pion field, which corresponds to quantum fluctuations round the hat, or the bottom of the bottle, feels no resistance and is massless. Of course, the $u$ and $d$ quarks are not quite massless and there is a small explicit breaking of chiral symmetry. Nevertheless, the pions remain by far the lightest of all hadrons, just as experiment requires.

What in the quark and gluon world drives this spontaneous chiral symmetry breaking in the hadronic sphere is a dynamical breakdown of the chiral symmetry of the quark fields. While the $u$ and $d$ quarks are nearly massless at short distance, the condensates in the vacuum provide a \lq\lq sticky medium'' through which these quarks propagate. Consequently, over longer distances of the size of a hadron, these quarks change from current to constituent quarks --- they become \lq\lq dressed''~[29]. 
\begin{figure}[t]
\begin{center}
\mbox{~\epsfig{file=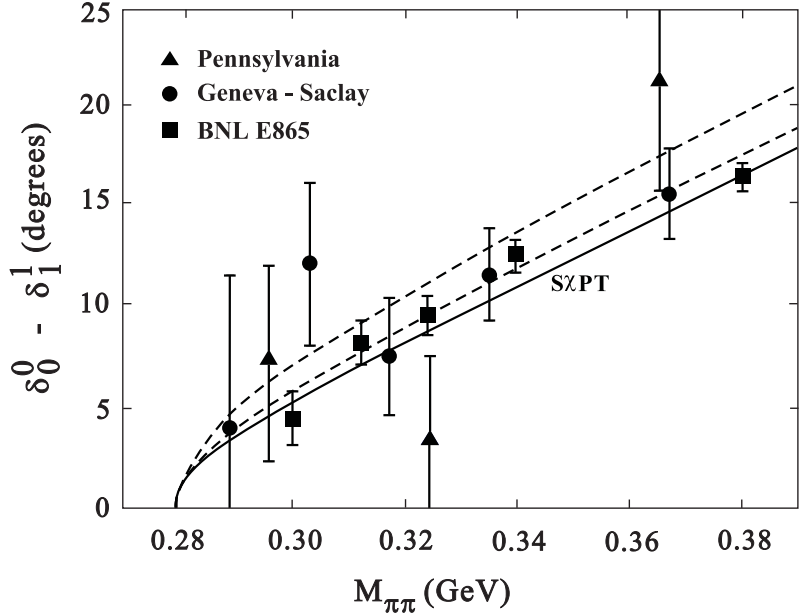,angle=0,width=10.5cm}}
\end{center}
\vspace{-3mm}
\caption{\leftskip 5mm\rightskip 5mm{The phase difference $\delta^0_0 - \delta^1_1$ as determined from $K_{e4}$ experiments --- from Pennsylvania~[33], Geneva-Saclay~[34] and BNL E865~[35]. The three curves with increasing phase difference  represent decreasing $\langle q {\overline q} \rangle$~[36]. The curve labelled S$\chi$PT is the prediction of Standard Chiral Perturbation Theory~[37].}}
\vspace{-2mm}
\end{figure}

Now what condensates drive this symmetry breaking~[27,30]? This, in fact, we can test from hadronic experiments. As a result of pions being the Goldstone bosons
of chiral symmetry breaking, the strong interaction of pions is forced to be \lq\lq weak'' at low energies. This requires the $\pi\pi$ scattering amplitude to have a zero in the near threshold region. Indeed, the amplitude has a line of zeros, which in the physical regions generate dips in the
corresponding differential cross-sections. The position of this zero contour depends crucially on the explicit breaking of chiral symmetry~[27,30,31]. If the $\langle q {\overline q}\rangle$ condensate is large, of the order of 250-270 MeV in scale, the zero in the $\pi^+\pi^-\to\pi^0\pi^0$ passes close to the symmetry point of the Mandelstam triangle, as required by Standard Chiral Perturbation Theory~[27]. If the  $\langle q {\overline q}\rangle$ condensate is smaller, and some other condensate drives chiral symmetry breaking,
 the line of zeros passes further from this point~[31].
 The shift in the zero contour is small, but nevertheless this can be checked by having precision data on $\pi\pi$ scattering in the very low energy region. Since pions are the lightest of all hadrons, their scattering is universal, being independent of their production process. Thus, for instance, by studying $K_{e4}$ decay, in which $K\to e\nu(\pi\pi)$, as a function of the 5 kinematic variables on which it depends~[32], we can extract the relative phase of the contributing $\pi\pi$ amplitudes. This phase difference is the same as that for $\pi\pi$ scattering itself.
 Thanks to the BNL-E865 experiment~[35], as shown in Fig.~2, we have a determination of the phase difference of sufficient precision to show that experiment is consistent with
a large $\langle q{\overline q} \rangle$ condensate saturating the Gell-Mann-Oakes-Renner relation~[38]
\begin{equation}
m_{\pi}^2\, f_{\pi}^2\;=\; - (m_u + m_d)\,\langle q{\overline q} \rangle\,+\,O(m_q^2)\quad .
\end{equation}
 to better than 90\%~[39]. Thus
neglecting higher order quark mass terms accords with $\langle q{\overline q} \rangle \simeq - (270 {\rm MeV})^3$ at a scale of 2 GeV: just as indicated by sum rule phenomenology, Table~2.

\section{Strong QCD}

\baselineskip=6.mm

In both the approach of effective field theory and that of QCD sum rules we regard the two worlds of hadron physics and quark dynamics as distinct. Clearly they are not: quarks create hadrons, hadrons are made of quarks. QCD must be able to explain the confinement aspects of quarks and gluons. Lattice Monte Carlo methods are often advertised as the way to bridge the gap between the two descriptions. This modelling is however quite unsuited to studying dynamical chiral symmetry breaking, since light quarks do not fit on a finite size lattice. Continuum approaches to QCD are much more appropriate to this problem~[40]. One considers the Schwinger-Dyson equations of QCD. These are an infinite set of nested integral equations, which contain all the information about the theory. These equations are most easily understood diagramatically. They say, for instance Fig.~3, that the complete quark propagator is determined by knowledge of the gluon propagator and the full quark-gluon interaction.
\begin{figure}[b]
\vspace{2mm}
\begin{center}
\mbox{~\epsfig{file=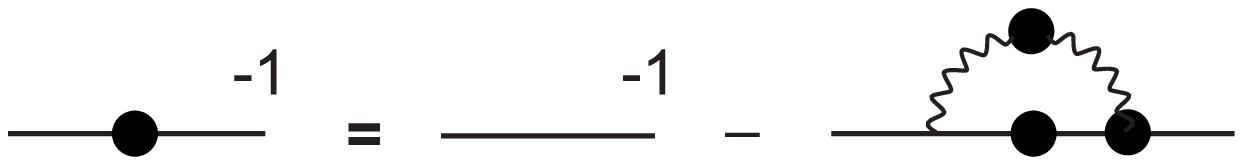,angle=0,width=13cm}}
\end{center}
\caption{\leftskip 5mm\rightskip 5mm{Schwinger-Dyson equation for the inverse of the \lq\lq dressed'' quark propagator showing how it is determined by the \lq\lq bare'' quark propagator plus dressing from a loop with the full gluon propagator and the complete quark-gluon interaction.}}
\vspace{-2mm}
\end{figure}
These in turn are determined in terms of higher point interactions. Consequently, the equation for the quark propagator can only be solved if the infinite set of equations is truncated. 

\begin{figure}[h]
\begin{center}
\mbox{~\epsfig{file=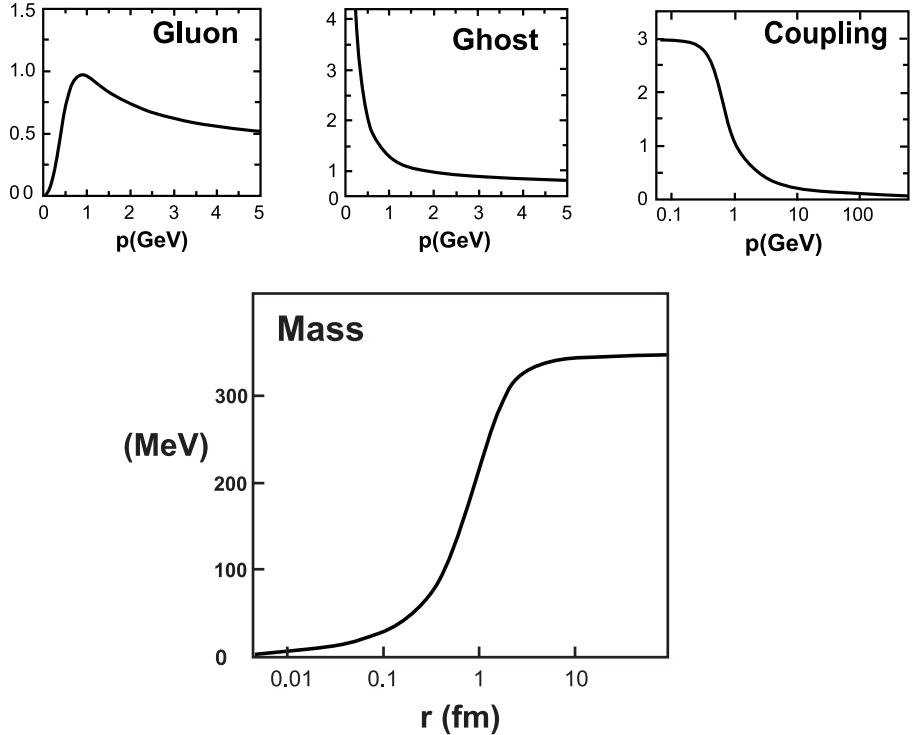,angle=0,width=12cm}}
\end{center}
\vspace{-1mm}
\caption{\leftskip 5mm\rightskip 5mm{Momentum dependence of the Landau gauge gluon and ghost dressing functions from  Schwinger-Dyson studies~[46]. These give the {\it effective} quark-gluon coupling shown. How the resulting $u$, $d$ quark mass function varies with distance of propagation is illustrated.}}
\vspace{-2mm}
\end{figure}

Perturbation
 theory, in which one makes an expansion of each interaction in powers of the coupling, is the best known truncation procedure and is in fact the only one that has been formally shown to respect gauge invariance and multiplicative renormalizability order by order. However, dynamical chiral symmetry breaking is definitely non-perturbative. It is a feature of strongly interacting theories and cannot occur at any finite order in perturbation theory.
Fortunately more relevant non-perturbative truncations have been developed over the past ten years~[41,40]. These enable the {\it up} and {\it down} quark propagators to be studied more precisely.  Solving the quark equation requires us to know about the behaviour of the gluon and ghost propagators and the quark-gluon interaction, Fig.~3. Twenty years ago it was believed the gluon propagator
in covariant gauges  was enhanced at small momenta~[42] and it was this that
was responsible for confinement. More recent work, largely by the T\"ubingen group~[43,44,45], has shown that in fact in the Landau gauge, the gluon propagator is suppressed at infrared momenta, while the ghost is strongly enhanced, Fig.~4.
Recall it is only a combination of gluon {\bf and} ghost propagators that can have physical degrees of freedom. These in turn make the effective quark-gluon coupling large at low momenta, as seen in Fig.~4. It has long been known that this is sufficient to generate dynamical chiral symmetry breaking. What is new is the detailed calculation that shows the $u$, $d$ quark mass, while being very small for large momenta, really does grow to 350 MeV in the low energy regime (Fig.~4), provided the \lq\lq right'' interaction is included~[46]. While the behaviour of the gluon, ghost and quark propagators is gauge dependent, the $\langle q {\overline q} \rangle$ condensate can be defined in a gauge invariant way. These calculations correspond to
\begin{equation}
\langle q{\overline q} \rangle\;\sim\; - (250-270\,{\rm MeV})^3
\end{equation}
as we might have hoped. This is most reassuring. 

The constituent mass of 350 MeV is reflected directly in the  masses of the scalar and vector mesons, for instance, and the mass of the nucleon. In contrast, the pion mass is uniquely tied to the current quark mass and is small.
While very small momenta are not accessible on the lattice, the gluon and ghost propagators found on the lattice at larger momenta~[47] agree with the continuum results just described. Consequently, undoubted progress is being made in extending the
predictions of QCD from the perturbative regime to confinement scales in a quantifiable way.

\section{Meson spectrum and decay} 
\baselineskip=6mm

A major success of Lattice Monte Carlo methods is the way it allows the study of
the gauge sector of QCD. Thus for years we have known that the world without quarks has a spectrum of colour singlet states of pure glue~[48]. Of these glueballs the scalar is the lightest with a mass of 1.5--1.7 GeV~[49]. 
There are scalar glueball candidates in this very mass region~[50,51]. Of course,
 these have been found in experiments \lq\lq with'' quarks. Glueball states have to decay to $\pi\pi$, $K{\overline K}$, $4\pi$, etc. Consequently, they must couple to quarks, Fig.~5, and so inevitably mix with other $q{\overline q}$ mesons with the same vacuum quantum numbers. This makes unambiguous identification of gluonic states far from easy. The isoscalar scalar sector is indeed complicated experimentally --- perhaps it had to be that way.
\begin{figure}[h]
\begin{center}
\mbox{~\epsfig{file=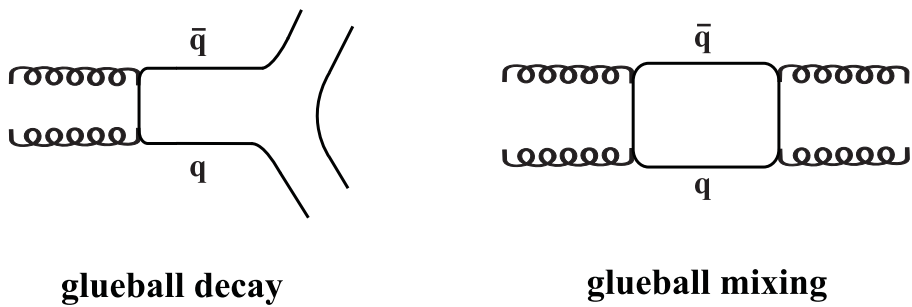,angle=0,width=12.cm}}
\end{center}
\vspace{-2mm}
\caption{\leftskip 5mm\rightskip 5mm{Perturbative picture of glueball decay and glueball mixing with $q{\overline q}$ mesons.}}
\vspace{-2mm}
\end{figure}

What we have learnt is that the states of the quark model are most easily identified with the hadrons we observe experimentally when {\it unquenching} is unimportant, Fig.~6. Thus the $\phi$ is readily seen to be an $s{\overline s}$ state and the $\rho$ and $\omega$ combinations of $u{\overline u}$ and $d{\overline d}$. This follows  from their respective decays to $K{\overline K}$, and to $2\pi$ and $3\pi$.
Though these decay modes are a crucial characteristic of their make-up, they have a relatively small effect on the states themselves. This is in part because of the $P-$wave nature of their hadronic {\it dressing}. This small effect  reproduces the suppression of the $1/N_c$ expansion. In contrast, $q{\overline q}$ scalar mesons are strongly disturbed by their couplings to open hadron channels~[51,52], Fig.~6. Thus almost
regardless of their composition in the {\it quenched} approximation, the $f_0(980)$ and $a_0(980)$ are intimately tied to the opening of the $K{\overline K}$ threshold. Scalars change on unquenching. For them the $1/N_c$ suppression of quark loops does not occur.  The fact that these resonances, $f_0(980)$ and $a_0(980)$, couple to both $\pi\pi$/$\pi\eta$ and $K{\overline K}$, means scalar non-strange and $s{\overline s}$ states communicate, Fig.~6. The coupling of different flavour quark pairs
is not merely unsuppressed, nullifying the OZI rule in the scalar sector~[54], but is even enhanced. This reflects the strange quark pairs in the vacuum~[55].  The flavour structure of the vacuum and how it changes between the world of two light flavours, $u$ and $d$, and the theoretically interesting limit in which the strange quark is also light (compared to $\Lambda_{QCD}$), is an open question~[56,57].
\begin{figure}[h]
\begin{center}
\mbox{~\epsfig{file=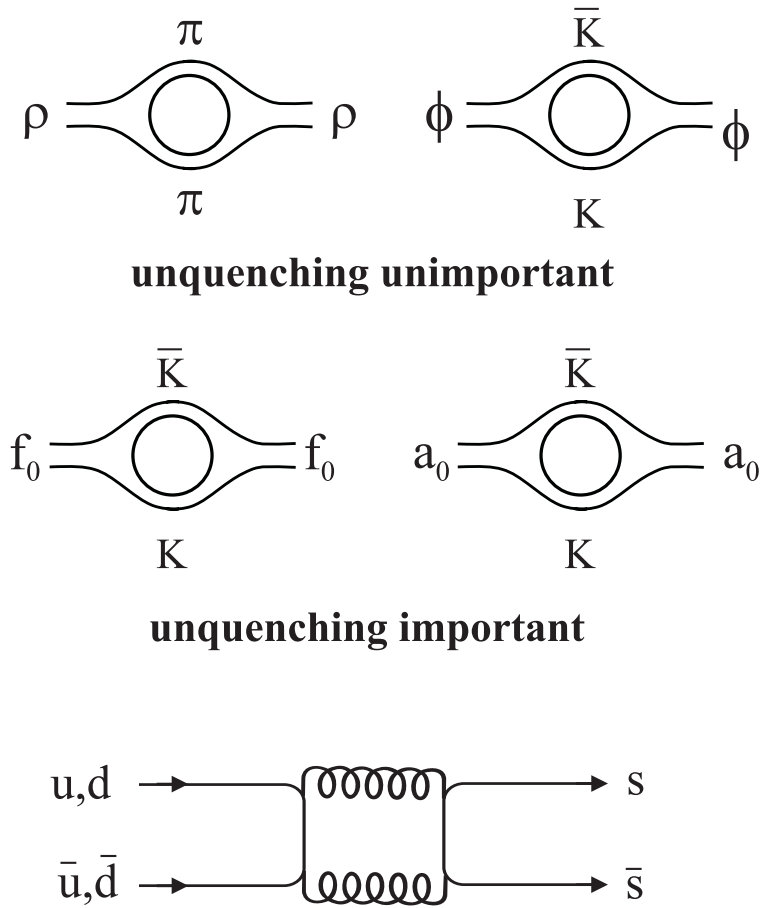,angle=0,width=10cm}}
\end{center}
\caption{\leftskip 5mm\rightskip 5mm{\lq\lq Unquenching'' of quark model states to make real hadrons has little effect on the vector mesons, $\rho$ and $\phi$, beyond allowing them to decay. In contrast the observed properties of the two scalar mesons $f_0(980)$ and $a_0(980)$ are produced by \lq\lq dressing''. These states then  enhance the coupling of $u{\overline u}$, $d{\overline d}$ systems to $s{\overline s}$ with 
no OZI suppression in scalar channels.}}
\end{figure}

The key role played by scalars means that
some or all of the scalar mesons, $f_0(400-1200)$, $f_0(980)$, $f_0(1370)$, $f_0(1500)$, $f_0(1710)$, ... , are closely related to the vacuum. 
They link the \lq\lq hadron  text'' to the {\it Book of QCD}. To decipher this relation we need more information. This is increasingly provided almost entirely from heavy flavour decays like
$\psi \to \phi X$~[58], $D_s \to \pi X$~[59], $\phi \to \gamma X$~[60]. The data are already  of excellent quality. We can learn a lot, once these experimental results are analysed in ways consistent with the complementary information collected over decades from purely hadronic reactions like peripheral and central $\pi\pi$ and  $K{\overline K}$ production~[61,62].

That meson physics is still interesting in 2002 is a consequence of the complex structure of the QCD vacuum  --- a complexity long recognised but now 
starting to be understood sufficiently well that it can be calculated.  
Hopefully this conference will add to this understanding.
\vspace{3mm}

\noindent{\bf Acknowledgements}

It is a pleasure to thank Andrzej Magiera and  colleagues 
for organising a most interesting meeting in such wonderful surroundings.
Partial support from the EU-TMR Programme, Contract No. CT98-0169, EuroDA$\Phi$NE is gratefully acknowledged.

\baselineskip=5.mm
\parskip=0.7mm

\end{document}